\documentclass[traditabstract]{aa} 
\usepackage{graphicx}
\usepackage{txfonts}
\usepackage{natbib}
\begin{document}

\authorrunning{R. Delgado-Serrano et al.}

   \title{How was the Hubble sequence 6 Gyrs ago?}

   \subtitle{}

   \author{R. Delgado-Serrano
          \inst{1,2}\fnmsep\thanks{Contact \email{rodney.delgado@obspm.fr}},
          F. Hammer
          \inst{1}\fnmsep\thanks{Contact \email{francois.hammer@obspm.fr}},
          Y. B. Yang
          \inst{1,3},
          M. Puech
          \inst{1},
          H. Flores
          \inst{1},
          \and
          M. Rodrigues
          \inst{1}
          }

   \institute{GEPI, Paris Observatory, CNRS, University of Paris-Diderot; 5 Place Jules Janssen, 92195 Meudon, France 
   \and  
            Technological University of Panama, 0819-07289 Panama, Rep. of Panama
   \and     
            National Astronomical Observatories, Chinese Academy of Sciences, 20A Datun Road, Chaoyang District, Beijing 100012, PR China
             }

   \date{submitted: 16 June 2009}

  \abstract
{The way galaxies assemble their mass to form the well-defined Hubble sequence is amongst the most debated topic in modern cosmology. One difficulty is to link distant galaxies, which emitted their light several Gyrs ago, to those at present epoch. Such a link is affected by the evolution or the transformation of galaxies, as well as by numerous selection and observational biases.
We aim to describe the galaxies of the Hubble sequence, 6 Gyrs ago. We intend to derive a past Hubble sequence that can be causally linked to the present-day one, and further estimate the uncertainties of that method. We selected samples of nearby galaxies from the SDSS and of distant galaxies from the GOODS survey. We verified that each sample is representative of galaxies selected by a single criterion, e.g., M$_{J}(AB)< -20.3$. We further showed that the observational conditions needed to retrieve their morphological classification are similar in an unbiased way. Morphological analysis was done in an identical way for all galaxies in the two samples. 
We found that our single criterion is particularly appropriate to relating distant and nearby galaxies, either if gas is transformed to stars in relatively isolated galaxies or, alternatively, if they accrete significant amounts of gas from the intergalactic medium. Subsequent mergers during the elapsed 6 Gyrs, as well as evolution of the stellar populations, are found to marginally affect the link between the past and the present Hubble sequence. Indeed, uncertainties from the above are below the errors due to the Poisson number statistics. We do find an absence of number evolution for  elliptical and lenticular galaxies, which strikingly contrasts with the strong evolution of spiral and peculiar galaxies. Spiral galaxies were 2.3 times less abundant in the past, which is compensated exactly by the strong decrease by a factor 5 of peculiar galaxies. It strongly suggests that more than half of the present-day spirals had peculiar morphologies, 6 Gyrs ago, and this has to be taken into account by any scenario of galactic disk evolution and formation. The past Hubble sequence can be used to test these scenarios and to test evolution of fundamental planes for spirals and bulges.}

   \keywords{  Galaxies: evolution --
                Galaxy: formation --
                Galaxies: spiral --
                Galaxies: peculiar --
                Galaxies: starburst --
                Galaxies: bulges --}

   \maketitle
%

\section{Introduction}
Galaxies are complex objects containing several tens of billions stars, as well as gas and dust. Present-day massive galaxies are regular and relaxed systems, and are made of a dispersion-supported bulge that is surrounded by a rotationally supported disk. They all fit the initial scheme proposed by \citet{1926ApJ....64..321H}, which is called the Hubble sequence. In the Hubble scheme, which is based on the optical appearance of galaxy images on photographic plates, galaxies are divided into three general classes: ellipticals (E), lenticulars (S0), and spirals. In the sequence, galaxies are organised from pure bulge (the elliptical) to increasing disk contribution to their light or mass (S0 to Sc). Further improvement has defined a fork that splits barred (SB) from non-barred spirals. There are also few massive galaxies, less than 10\%, showing irregular or peculiar morphologies, and these escape to this classification. 

How was the Hubble sequence in the past? The Hubble Space Telescope (HST) revealed the morphologies of distant galaxies, and a simple inspection of the deepest observed fields shows how galaxy morphologies have evolved \citep{2002PASP..114..797V}. Indeed, we do not lack morphological details as the HST/ACS instrument allows investigation of the details of 200 pc at $z\sim$ 1. However, this does not suffice to establish the past Hubble sequence. First, the cosmological dimming is so severe that only long exposures, such as in GOODS or UDF, can detect the optical radius of a Milky Way like galaxy at $z$=0.4-0.5 and 1.3-1.5, respectively.  Second, the past Hubble sequence has to be made of galaxies that can be connected with the present-day ones: sporadic events such as galaxy collisions could have dramatically affected this link. Third, although distant galaxies are linked to present-day ones through the isotropic principle, it needs a simple (and preferentially single) criterion in selecting them in two different redshift slices.

The motivation of this study is to establish the Hubble sequence in the past and to verify to what accuracy it can be linked to the sequense at the present epoch. A considerable advantage would be to compare local galaxies to those at moderate redshifts, in order to limit most of the complexities discussed above. At $z$=0.65, corresponding to an epoch of 6 Gyrs ago, the GOODS imagery -after correction for cosmological dimming-  is as deep as or deeper than the SDSS imagery at $z$=0. Moreover, it is unlikely that major mergers could significantly affect the number density of galaxies since that epoch. The selection criterion has to be kept simple and should address the mass in one way or another. There is a relatively good correspondence between stellar mass and absolute magnitude in the near-IR, so we choose a limit $M_{J}(AB)$ $< -$20.3. Such a value is adopted from the IMAGES\footnote{Intermediate MAss Galaxy Evolution Sequence is a VLT large program aiming at recovering the spatially resolved kinematics of $\sim$ 0.4$< z <$ 0.9 galaxies, see \citet{2008A&A...477..789Y}.} survey and corresponds approximately to $M_{stellar}$ $>$ 1.5 $10^{10}$ $M_{\odot}$. However, one needs to make it clear whether this simple criterion select similar galaxy masses in the two redshift bins.

The methodology for morphological classification is a crucial issue, especially when comparing galaxies at two different epochs. Quoting \citet{2004ARA&A..42..603K}, ``it is useful only if classification bins at least partly isolate unique physics or order galaxies by physically relevant parameters. The Hubble-Sandage-de Vaucouleurs classification scheme has done these things remarkably well". We thus adopt a methodology \citep[see, e.g.,][]{2008A&A...484..159N} that systematically compares morphologies of distant galaxies to those of local galaxies. Identification of peculiarities would thus be directly assessed by the discrepancy between a given galaxy and the galaxies that populate the local Hubble sequence. Such a technique is based on a simple and reproducible decision tree and a full decomposition of the galaxy light profile in two components, namely the bulge and the disk. A striking result of \citet{2008A&A...484..159N} is the excellent agreement between such a morphological classification and classification of their kinematics. By studying a representative sample of 52 distant galaxies, they demonstrate that 80\% of rotating galaxies possess spiral-like morphologies, and that 95\% of galaxies with complex or chaotic kinematics have peculiar morphologies.  In other words, peculiar morphologies are associated with anomalous kinematics. \citet{2008A&A...484..159N} has also shown that other classifications based on automatic assessment of  compactness or asymmetry do not correctly predict the kinematics, as they overestimate the number of spiral-like morphologies. Such an effect has indeed been noticed by \citet{2003AJ....126.1183C,2005ApJ...620..564C} \citep[see also][]{2007ApJ...660L..35K}.

Finally, when comparing galaxy morphology at different redshifts, one must pay attention to all possible biases linked to the use of different instruments and, more importantly, must minimise morphological $k$-corrections. In the present paper, after considering the above effects, we generalized the morphological classification method from \citet{2008A&A...484..159N} and we applied it to a local sample of galaxies observed by SDSS and to a distant sample of galaxies observed by HST/ACS, in order to compare their properties. 

Samples and their selection are described in Sect. 2. Methodology and data analyses are shown in Sect. 3. Results of our morphological classification are presented in Sect. 4, and Sect. 5 discusses to which accuracy one may establish the past Hubble sequence. Finally, in Sect. 6 we summarize our conclusions. Throughout this paper we adopt cosmological parameters with $H_{0}= 70 km s^{-1} Mpc^{-1}$, $\Omega_{\Lambda}$ = 0.7. Magnitudes are given in the AB system.

\section{Representative samples of nearby and distant galaxies}

The first sample represents the local galaxies, constructed from SDSS survey. The second one represents the distant galaxies selected from GOODS.  To gather both samples, we used a single selection criterion, i.e., the absolute magnitude in J band (M$_{J}<-$20.3). However for practical reasons we further restricted our sample to the galaxies possessing a good quality spectrum to measure {\sc [Oii]}$\lambda$3727 emission line, as well as  having high resolution images in at least three optical bands. These further limitations are only rejecting objects for technical reasons\footnote{This include bad quality imagery and bad pixels near the {\sc [Oii]}$\lambda$3727 emission line.} in the local Universe since data from SDSS have reached an incomparable homogeneity. While the GOODS image quality and depth are as well homogenous for distant galaxies, spectroscopic surveys show quite significant heterogeneities. We followed the steps of \citet{2007A&A...465.1099R} showing the ability in gathering a representative sample from the collection of spectroscopic samples in the GOODS area. All in all, the luminosity functions of both samples are compared to those of similar surveys, and an excellent agreement is found.

\subsection{Local galaxy sample}

Galaxies were selected from the \citet{2007AJ....134..579F} sample. It includes 2253 galaxies with Petrosian magnitude lower than r$_{P}$=16, that are lying in a 230 deg$^{2}$ rectangular area in the equatorial area of the northern sky. From this sample we only kept galaxies with available spectra and imaging from the Sloan Digital Sky Survey \citep[SDSS;][]{2000AJ....120.1579Y} Data Release three \citep[DR3;][]{2005AJ....129.1755A}. All the objects have images in u $($3551 $\mathring{A})$, g $($4686 $\mathring{A})$, r $($6165 $\mathring{A})$, i $($7481 $\mathring{A})$ and z $($8931 $\mathring{A})$ bands.  Relative magnitudes were retrieved from the 2MASS \citep{2006AJ....131.1163S} and SDSS-DR3 catalogs.  It results a selection of 2113 galaxies. We then considered only the 1665 galaxies with an absolute magnitude M$_{J}<-$20.3, following our selection criterion. Figure 1 shows the M$_{J}$ vs redshift plane of the local SDSS catalog from \citet{2007AJ....134..579F}. There is a clear lack of intrinsically faint galaxies at $z >$ 0.03, especially blue galaxies with magnitude just below  r$_{P}$=16. This is illustrated by the blue curve which corresponds to a flat spectrum between r-band and J-band, i.e., that of a pure starburst at the sample magnitude limit. Figure 1 shows that at $z <$ 0.03, all M$_{J}<-$20.3 are included in the \citet{2007AJ....134..579F} sample. It represents a sample of 218 galaxies (red points). Finally, from these 218 galaxies, we only kept those whose spectra  include the {\sc [Oii]}$\lambda$3727 emission line, further reducing the sample to 116 galaxies.  These galaxies have a redshift of 0.0207$\leqslant z \leqslant$0.030. 

Figure 2$($a$)$ compares the  distribution of the J-band absolute magnitudes of these galaxies with the J-band luminosity function of local galaxies \citep{2006MNRAS.369...25J}. A Kolmogorov-Smirnov test indicates a probability of 98$\%$ that this sample and the local luminosity function have a similar distribution. Thus our local sample of galaxies is representative of the M$_{J}<-$20.3 galaxies in the local Universe. Additional information about this sample observed by the SDSS can be found in \citet{2007AJ....134..579F} and references therein.

\subsection{Distant galaxy sample}

This sample is made up of the combination of two subsamples, distant star-forming galaxies and distant quiescent galaxies. Those have been arbitrarily defined by their {\sc [Oii]}$\lambda$3727 EW, larger or smaller than 15$\mathring{A}$, respectively \citep[see][]{1997ApJ...481...49H}. In both cases, galaxies were selected with the following criteria: M$_{J}(AB)< -20.3$ and redshift range $0.4<z<0.8$. All these limits were adopted to be consistent with the  IMAGES survey. Furthermore, we selected only galaxies having GOODS(ACS) v2.0 images in at least 3 bands $($v $-$ 5915 $\mathring{A}$, i $-$ 7697 $\mathring{A}$ and z $-$ 9103 $\mathring{A})$.

There is a relatively good correspondence between star-forming and quiescent galaxies defined as above with blue and red galaxies, respectively. Figure 3 compares rest-frame (U-B) colors of galaxies in the distant sample with the limit applied by \citet{2007MNRAS.380..585C} to separate blue from red galaxies. It is, however,  slightly imbalanced as there are more quiescent galaxies in the blue cloud than star-forming ones in the red cloud. An optimal balance would be provided if we had chosen an {\sc [Oii]}$\lambda$3727 EW of  about 11$\mathring{A}$. This might slightly alter the comparison with \citet{2007MNRAS.380..585C} in Fig.~2 (b, c, see also below).

\paragraph{Distant star-forming galaxies:}

This sample of 52 starburst galaxies (EW({\sc [Oii]}$\lambda3727) \geqslant 15 \mathring{A})$ has been gathered by \citet{2008A&A...484..159N} in the frame of the IMAGES survey. We kept only 49 galaxies, and two were rejected because they do not have at least 3 bands GOODS(ACS) v2.0 images $($J030238.74+000611.5 $\&$ J223256.08-603414.1$)$ and one of them (J033210.76-274234.6) has been recently identify as a quiescent galaxy \citep{2009A&A...501..437Y}.  Figure 2(b) shows the distribution of the J-band absolute magnitudes of the distant star-forming galaxy subsample. To further verify its representativeness, we over-plotted in this histogram the K-band luminosity function obtained by \citet{2007MNRAS.380..585C} for blue galaxies in the redshift range $0.25<z<0.75$. The correction from K$-$band to J$-$band is very small in the AB system, according to the available values calculated in the IMAGES survey. The J-K values average to -0.08$\pm$0.03 and are almost independent of the spectrophotometric type \citep[see also][]{2001MNRAS.326..745M}. In the following we assume a J-K correction of -0.1 mag. Our sample follows the same distribution of the  luminosity function (Kolmogorov-Smirnov test, with a probability of 97$\%$), as illustrated in Fig.~2(b). It confirms the \citet{2008A&A...477..789Y} and \citet{2008A&A...484..159N} results that the IMAGES survey is representative of distant starbursts.

\paragraph{Distant quiescent galaxies:}

This subsample complements the previous subsample, as it includes only galaxies with (EW({\sc [Oii]}$\lambda3727)< 15 \mathring{A})$. It was selected in the GOODS area, using spectroscopic data collected from (1) observations made by VLT/FORS2 by IMAGES \citep{2008A&A...492..371R}, (2) FORS2 observation in the frame of ESO follow-ups \citep{2005A&A...434...53V,2006A&A...454..423V}, (3) observations made by VLT/VIMOS \citep{2005A&A...439..845L}. For the VVDS survey, only secure redshift were used (with zflag$=3/4$), which correspond to more than 95$\%$ of confidence. The first step was to detect those spectra with a wavelength coverage including the {\sc [Oii]}$\lambda$3727 line (the same criterion that is applied to the local sample galaxies). We then took only those galaxies with EW({\sc [Oii]}$\lambda3727)< 15 \mathring{A})$ and an absolute magnitude M$_{J}<-$20.3 within the redshift range $0.4<z<0.8$ (same criteria of the IMAGES large program).
  Finally, we kept only the galaxies having GOODS(ACS) v2.0 images in at least 3 bands, leading to a sample of 94 quiescent galaxies. A high fraction $(77\%$ of the sample$)$ have spectra observed with VLT/FORS2 ensuring the best accuracy in measuring {\sc [Oii]}$\lambda$3727 EW. Figure 2(c) shows the corresponding distribution of the J-band absolute magnitudes that is compared to the luminosity function from \citet{2007MNRAS.380..585C} for red galaxies in the redshift range $0.25<z<0.75$. The J-K magnitude correction was done in the same way as in the previous subsample. A Kolmogorov-Smirnov test provides a probability of 94$\%$, that the quiescent galaxies sample and the corresponding luminosity function in the redshift range $0.25<z<0.75$ are drawn from the same distribution. Both distributions show well the peak that is due to massive early type galaxies.

   \begin{figure}
   \centering
   \includegraphics[width=8cm]{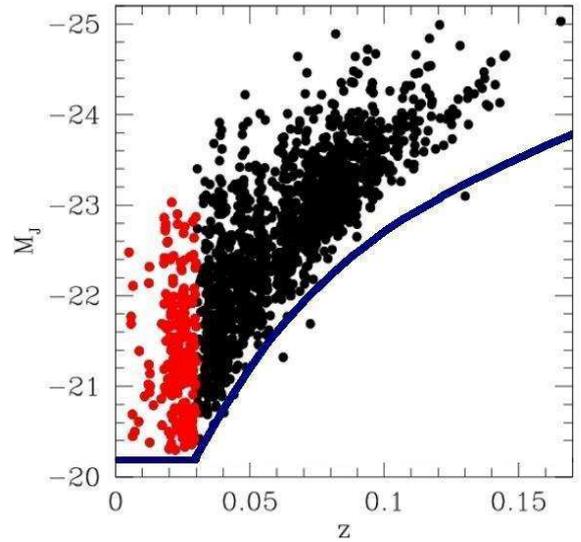}
      \caption{Galaxies of the \citet{2007AJ....134..579F} sample in the $M_{J}$ versus $z$ plane. The red points are those included in the selected representative sub-sample of 218 galaxies with $0.02<z<0.03$. This cut in redshift was determined to optimize the representativity in terms of the LF. Curved blue line shows the limit of the sample that is assumed to be represented by a blue starburst with r$_{P}$=16.}
         \label{malmquistLocal}
   \end{figure}

\begin{figure*}
   \centering
   \includegraphics[width=6cm]{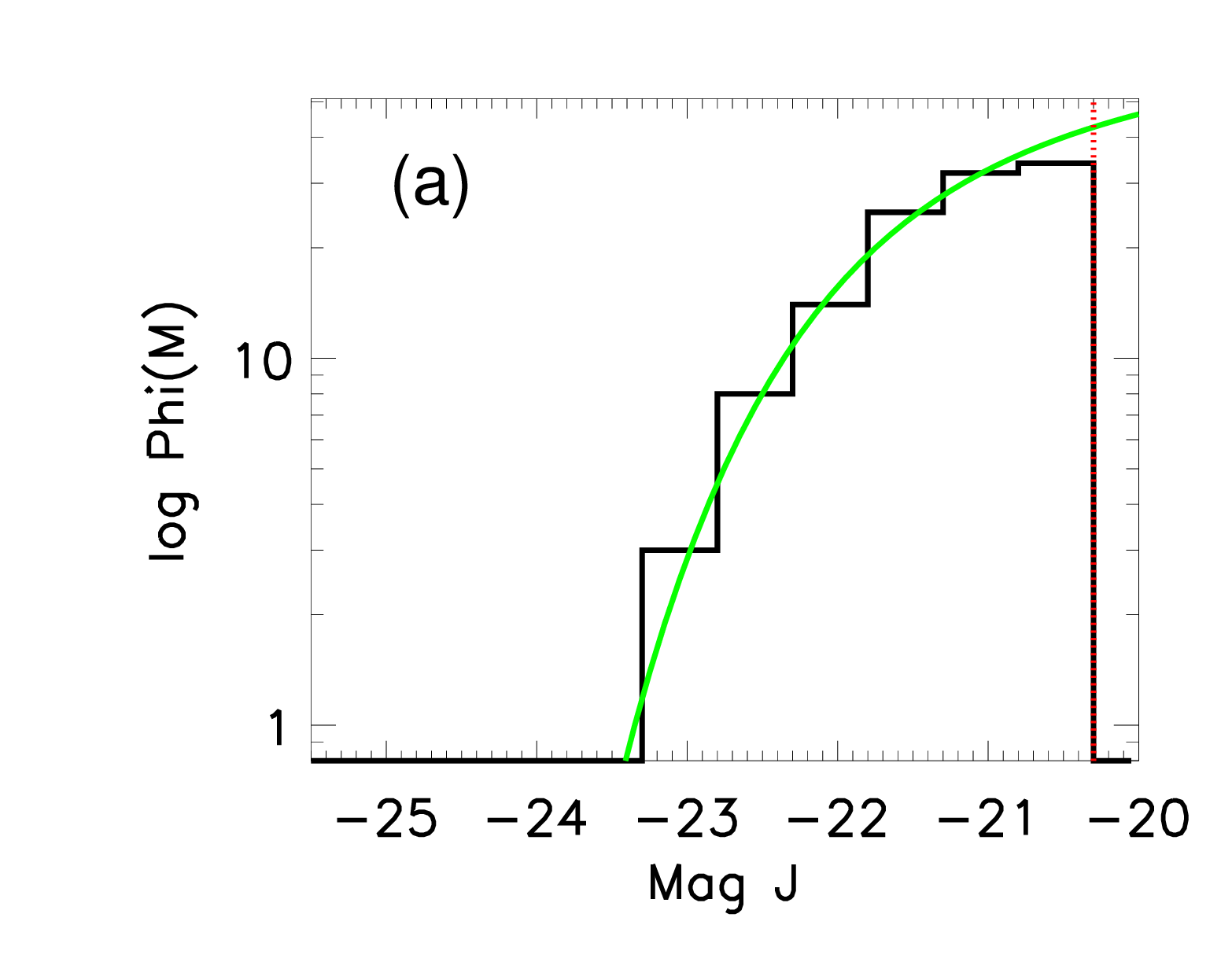}
   \includegraphics[width=6cm]{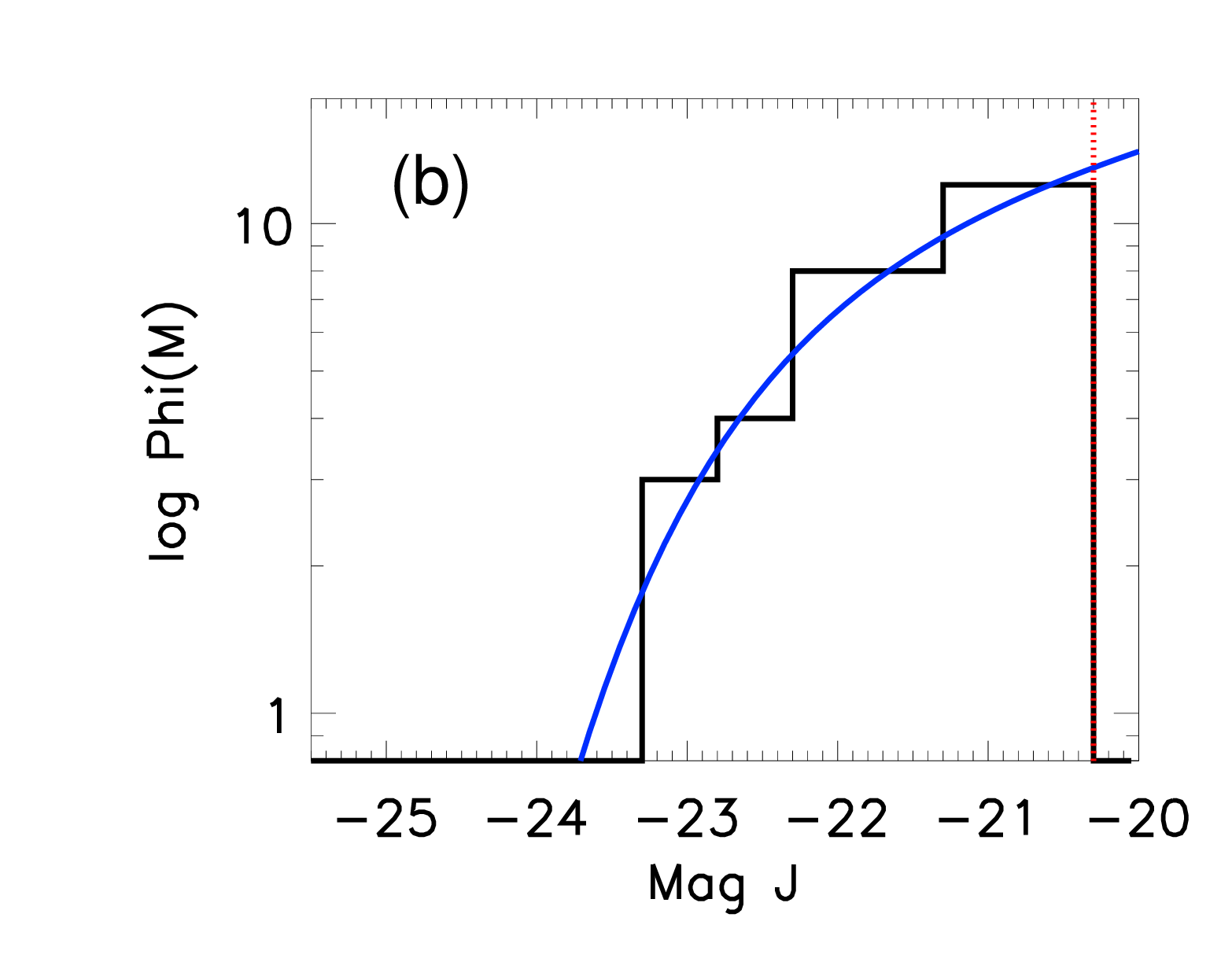}
   \includegraphics[width=6cm]{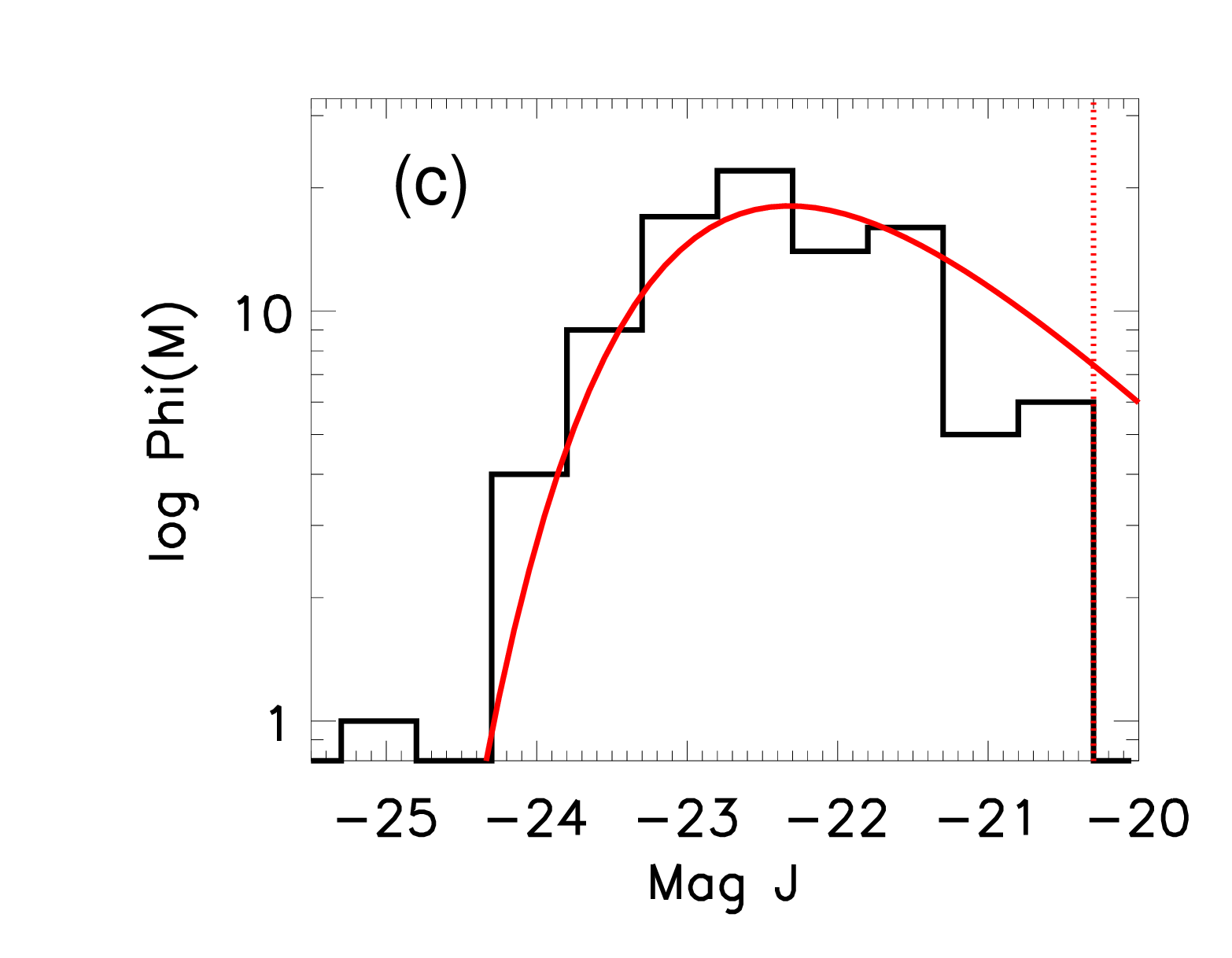}
   \caption{(a) Local galaxy sample. The black line shows the histogram 
                of $M_{J}$ for all 116 galaxies. Observed LF in the redshift range 
                $0.0025<z<0.2$ (green line) 
                is also plotted as a reference. This 
                observed LF come from \citet{2006MNRAS.369...25J}.
            (b) Distant star-forming galaxy sample. The black line shows 
               the histogram of $M_{J}$ for all 49 galaxies. Observed 
               LF for $0.25<z<0.75$ (blue line) is also plotted as a 
               reference. This observed LF comes from \citet{2007MNRAS.380..585C} 
               for blue galaxies.
            (c) Distant quiescent galaxy sample. The black line shows 
               the histogram of $M_{J}$ for all 94 galaxies. Observed LF 
               for $0.25<z<0.75$ (straight red line) is also plotted as a 
               reference. This observed LF comes from \citet{2007MNRAS.380..585C} 
               for red galaxies.
             In all three plots, the vertical dashed red line indicates the $M_{J}$ IMAGES limit of $-$20.3.  
            }
              \label{FigRep}%
    \end{figure*}
%

   \begin{figure}
   \centering
   \includegraphics[width=8cm]{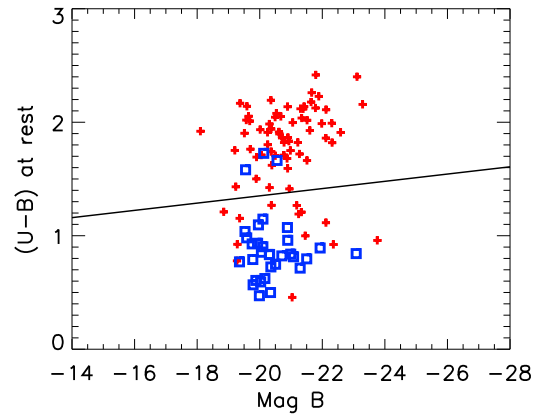}
   \caption{U-B color distribution of the distant sample (blue squares: starbursts, red crosses: quiescent galaxies). The black line represents the color-magnitude relation applied by \citet{2007MNRAS.380..585C} to separate their blue and red populations (see their Eq. 2).}
         \label{UBplots}
   \end{figure}
%

\section{A similar morphological analysis for both local and distant galaxies}

\subsection{Comparing the available data for the two samples}

Our main goal is to ensure that the comparison of morphologies is not affected by biases related to spatial sampling, depth, and $k$-corrections. In this respect, the detailed study of \citet{2008ApJS..175..105B} is very encouraging. They indeed created a set of SDSS galaxies that they have artificially redshifted to redshifts 0.1 $< z <$ 1.1, in order to compare them with GEMS and COSMOS galaxies. 

\citet{2008ApJS..175..105B} applied and verified whether these templates may be compared to galaxies observed with HST/ACS. They did not find any significant difference in estimating absolute magnitude, half light radius, and S\'{e}rsic index (see their Fig. 9). As they concluded, their software has no unwanted systematic effects on measured morphological and photometrical parameters of simulated galaxies, and deviations from theoretical values are well within the uncertainties expected from their GALFIT \citep{2002AJ....124..266P} simulations.

At $z$=0.65, the concordance of the sampling in physical units is optimal: the HST/ACS of GOODS delivers an FWHM of 0.108 arcsec that corresponds to 0.81 kpc.  At $z$=0.025 an SDSS galaxy is observed within an average FWHM=1.4 arcsec that corresponds to 0.74 kpc.  Both pixel sizes correspond to a sampling of the PSF by 3.6 and 3.5 pixels, respectively. Thus the conditions reproduced by \citet{2008ApJS..175..105B} are particularily appropriate for a comparative analysis of the local and distant samples of galaxies considered in this study. 

We thus have to consider two additional effects, i.e., the comparison of the optical depth of the two databases and the $k$-correction effects. Concerning the latter, there is excellent agreement between the rest-frame band of distant galaxies to those of local galaxies, assuming $z$=0.65 for the median value of distant galaxies (see Table 1).
Table 2 summarizes the observational conditions resulting from both SDSS and GOODS. We are interested in verifying whether the analysis of distant galaxies is based on data as deep as or deeper than those used for local galaxies:

   \begin{eqnarray}
       \frac{SNR^{HST}}{SNR^{SDSS}} = \sqrt{\frac{FWHM^{HST}}{FWHM^{SDSS}}}*\sqrt{\frac{T^{HST}}{T^{SDSS}}}*\frac{D^{HST}}{D^{SDSS}}  \nonumber \\
         \qquad  *\sqrt{\frac{B^{SDSS}}{B^{HST}}}*\frac{f_{\lambda}^{HST_{z=0.0}}}{f_{\lambda}^{SDSS}}*\frac{1}{(1+z)^{5}} 
    \end{eqnarray}

It turns out that, after accounting for cosmological dimming and $k$-correction (the last term in Eq.~1), the GOODS imaging is deeper than that of SDSS, by 0.52, 0.08, and 1.02 magnitude, in rest-frame u, g, and r bands, respectively. Possible effects related to depth will be discussed later in the discussion section. Besides this effect, we confirm the conclusion of \citet{2008ApJS..175..105B} that there should not be systematic effects on measuring morphological and photometrical parameters from one sample to the next.

%

\begin{table}
\label{SDSS_ACS_lambda}      
\centering                          
\tiny{
\begin{tabular}{c|c|c|c|c|c|c}        
\hline \hline  \bf{Survey}& \bf{--} & \bf{u band} & \bf{g band}& \bf{r band}& \bf{i band}& \bf{z band} \\ \hline
SDSS&--&3551 $\mathring{A}$&4686 $\mathring{A}$&6165 $\mathring{A}$&7481 $\mathring{A}$&8931 $\mathring{A}$ \\ \hline
\bf{--}& \bf{B band} & \bf{V band}& \bf{i band}& \bf{z band}& \bf{--}& \bf{--} \\ \hline 
GOODS&4312 $\mathring{A}$&5915 $\mathring{A}$&7697 $\mathring{A}$&9103 $\mathring{A}$&--&-- \\ \hline 
{\it rest-frame}&2582 $\mathring{A}$&3542 $\mathring{A}$&4609 $\mathring{A}$&5451 $\mathring{A}$&--&-- \\ \hline \hline
\end{tabular}
}
\caption{Wavelength comparison of different bands in SDSS and GOODS(ACS).}  
\end{table}


\begin{table*}
\label{SDSS_ACS_compa}      
\centering                          
\begin{tabular}{l | c c c | c c c c}        
\hline \hline  & \multicolumn{3}{c|}{SDSS} & \multicolumn{4}{c}{GOODS ACS}   \\ \hline 	
D=telescope diameter (m) &\multicolumn{3}{c|}{2.5} &\multicolumn{3}{c}{2.4} \\ \hline
Band & {\bf{u}}   & {\bf{g}}  & {\bf{r}} & {\bf{B}}  & {\bf{V}}  & {\bf{i}} & {\bf{z}}    \\ \hline
T=Expo-time (s) &53.907456 &53.907456 &53.907456 & 7200.00&5450.00 &7028.00 &18232.00   \\ 
B=sky background (mag) & 22.15&21.85 & 20.85&23.43 & 22.74&22.72 &22.36   \\ 
Filter FWHM ($\mathring{A}$) & 567.00&1387.00 & 1373.00&728.95 &1565.50 &1017.40 &1269.10   \\ 
Filter range ($\mathring{A}$) & $\thicksim$1000.00&$\thicksim$1800.00 &$\thicksim$1500.00 &8780.00 & 2570.00&1910.00 &$>$3080.00   \\  \hline \hline 
\end{tabular}
\caption{Observational parameters for both SDSS and GOODS(ACS) imaging.}
\end{table*}
%

\subsection{Light profile analysis}

Half-light radius and bulge-to-total ratio (B/T) are derived from the surface brightness profile analysis. In this work, instead of fitting simultaneously all neighboring objects within 5.0 arcsec of the target, they were masked to avoid any flux contamination. To measure the half-light radius, we have developed our own IDL procedure, which allows us to visually analyze different profiles of the galaxy at the same time: (1) the flux profile, (2) the magnitude profile, and (3) the flux content within ellipses of different radius with a step of one pixel. The first two let us determine a reasonable sky value and the last one gives us the half light-radius when a real ``plateau" is visualized. 

Each galaxy is decomposed as a combination of a bulge and a disk. This decomposition is a two-dimensional modeling based on light (flux) distributions. We then modeled each galaxy light profile with a combination of two S\'{e}rsic laws. The first one would represent the bulge with a S\'{e}rsic index that is left free, and the second one would represent the disk with a S\'{e}rsic index equal to one. B/T is derived from the GALFIT decomposition of the galaxy. 

Half-light radius measurements and GALFIT simulations were done at rest frame r band, i.e., at z band for galaxies in the distant sample.

\subsection{Color images and maps }

Color is an important information in our morphological classification process. We have two samples of galaxies (local and distant samples) and we want to compare them. It is needed to use same rest-frame colors in both samples to construct color maps (see Table 1). Therefore, in the case of two-color maps, we use u-r bands for the local sample and v-z bands for the distant sample. We do the same for three-bands color images (u-g-r and v-i-z bands, respectively).

Three-band color images are useful for the morphological classification for examining the small-scale structures of the galaxy. However, we need color maps in which color of individual features can be measured. To do so, we substract, pixel by pixel, the magnitude in two observed bands using an algorithm that allows estimation of colors and their uncertainties \citep[see details in][]{2005A&A...435..507Z}. A color map of any substructure in the galaxy can then be compared with models of stellar population synthesis. They can thus be useful for estimating the color of a bar or identifying dusty or star-forming regions.

   \begin{figure*}
   \centering
   \includegraphics[width=13cm]{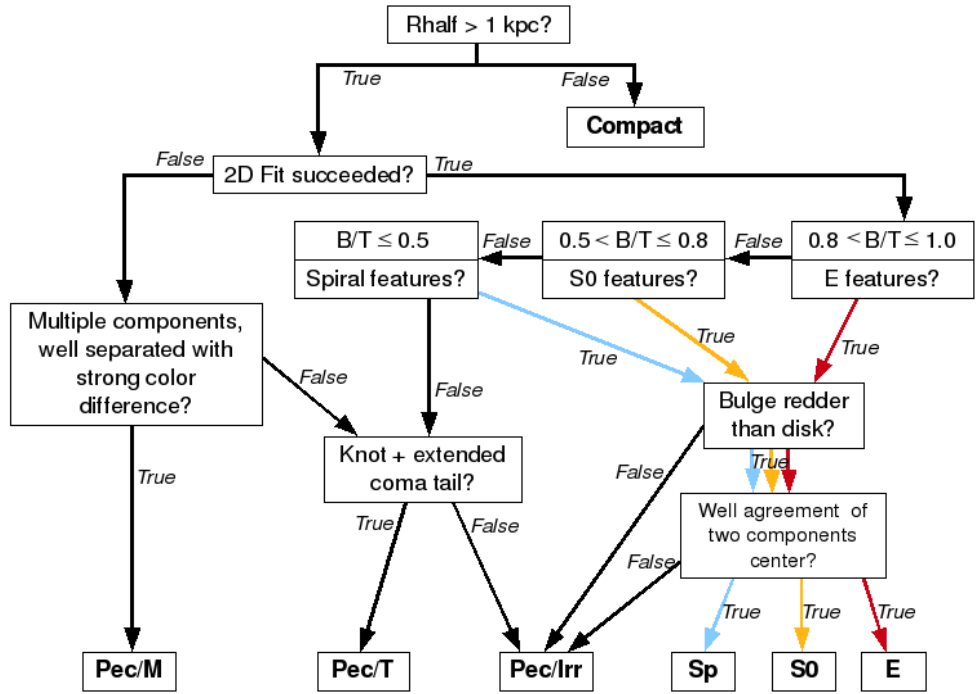}
   \caption{Semi-automatic decision tree used in the morphological classification process. Each step considers a simple and unique criterion. We chose a very conservative method, which takes into account the well known morphologies of local galaxies that populate the Hubble sequence.}
    \label{FigTree}
    \end{figure*}
%

\subsection{Morphological classification}

To make the morphological classification a reproducible process and to reduce its inherent subjectivity, we constructed a pseudo-automatic decision tree (described in Fig. 4). It was adapted from \citet{2008A&A...484..159N} to account for S0 and E galaxies. The decision tree takes all the available quantities into account such as B/T, half light radius, GALFIT parameters, GALFIT model and residual images, error images, disk-bulge-galaxy profile, two color-maps, and three-color images. To reduce the residual subjectivity, the morphological classification of each galaxy of the samples was done by three of us (RD, FH, YY) following the decision tree. Then, our classification was compared. Because of the relative simplicity of the decision tree, the agreement between the various classifications was excellent for both samples of local and distant galaxies.
\\

A few changes to the \citet{2008A&A...484..159N} methodology were made in the present morphological classification. First, we now formally account for the agreement between bulge and disk centers. Indeed, we have noticed that, in some cases, the fit bulge+disk of galaxies seems to be good, but in the three-color image or in the residual fit image we see disagreement between bulge and disk centers. Galaxies with a disagreement in their centroid larger than 3 pixels are considered as peculiar galaxies (see decision tree). Second, and conversely to what was chosen in \citet{2008A&A...484..159N}, we classify as compact all objects with a half-light radius below 1 kpc. In previous studies of $z \thicksim$0.6 galaxies, the limit to defining compact galaxies has been taken to be 3 kpc \citep{2005ApJ...632L..65M,2005A&A...435..507Z,2001ApJ...550..570H} because in all those works the spatial resolution of WFPC2 images was significantly higher than the 0".108 FWHM provided by  ACS/HST images for most (90$\%$) galaxies in the distant sample. We then have access to a better spatial resolution. Thus, the compactness limit is 1 kpc, and only such and extreme compactness is limiting our morphological classification.

To summarize, we distinguish between the following classes:

 \begin{enumerate}
      \item E: elliptical galaxies, with a B/T ratio between 0.8 and 1.0.
      \item S0: lenticular galaxies. There is no presence of regular structures (arms) and the bulge is redder than the disk. A highly symmetric disk, as well as the match of bulge and disk center, is observed. The B/T ratio has a value between 0.5 and 0.8.
      \item Sp: spiral disk galaxies. This class is characterized by a bulge redder than the disk. Regular arms and a highly symmetric disk are also present. Bulge and disk centers coincide, and the B/T ratio is smaller than 0.5. 
      \item Pec: peculiar galaxies. The main characteristic of this class is the presence of asymmetrical features. However, we may divide this class into four subclasses: possible mergers (Pec/M), which are objects with irregularities that could be associated to merger/interaction events; tadpole-like (Pec/T), which are objects showing a knot at one end plus an extended tail; irregulars (Pec/Irr) for objects with asymmetric irregularities that cannot be associated with arms, objects with a center (bulge) bluer than the rest, or objects with a disagreement between bulge and disk centers larger than 3 pixels; and compact (Pec/C), which are galaxies with a half-light radius smaller than 1.0 kpc. 
   \end{enumerate}   

According to this decision tree, we have classified the 259 galaxies that belong to our two samples. Galaxies from \citet{2008A&A...484..159N} have kept their morphological classification, except for those classified as compact by them but that are not in the present study.

\section{Morphological classification: results}

Table 3 summarizes the fraction of E, S0, spirals, and peculiar galaxies. Peculiar galaxies are also subdivided into 4 subcategories as described in Sect. 3.4.

\subsection{Morphological classification of local sample}

By applying our morphological classification method to the local sample, we confirm that the local Universe is dominated by spirals galaxies that represent 72$\%$ of the population, and that the fraction of E/S0 is 18\%, while 10$\%$  of galaxies have peculiar morphologies. This confirms the fraction of spirals in local galaxies as found by \citet{2005A&A...430..115H} on the basis of the morphological determined luminosity function of \citet{2004AJ....127.2511N}. The number of compact galaxies in the local sample is very small (2$\%$), which proves that spatial resolution can not significantly alter our classification. Among the 116 galaxies of the local sample we also find 96 (83$\%$) quiescent galaxies (EW({\sc [Oii]}$\lambda3727]) < 15 \mathring{A})$ and 20 (17$\%$) starburst galaxies (EW({\sc [Oii]}$\lambda3727) \geqslant 15 \mathring{A})$. \citet{1997ApJ...481...49H}, using a completely different method, find the same percentage of starburst galaxies which have same definition than in the present paper (see their Fig. 3 and their Sect. 3.1).

\subsection{Morphological classification of distant sample}

As assumed by \citet{2005A&A...430..115H} or by \citet{2008A&A...484..159N}, one would have expected a small fraction of peculiar morphologies in the sample of quiescent galaxies. In contrast to that expectation, we find a quite surprisingly high fraction of peculiar galaxies, 25$\%$. Peculiar galaxies are also the more common type (69$\%$) in the starburst distant sample, in which no elliptical or lenticular galaxies are found. Gathering the whole sample of distant galaxies, we find that 52\% of them have peculiar morphologies, while only 31\% have spiral morphologies, with the rest either E or S0. Gathering distant starbursts and quiescent galaxies in a single representative sample was done following \citet{1997ApJ...481...49H}: galaxies with significant line emission (EW({\sc [Oii]}$\lambda3727) > 15 \mathring{A})$ represent 60$\%$ of intermediate redshift  $(0.4<z<1.0)$ galaxies, while quiescent galaxies represent 40$\%$. Such a distribution is fairly well confirmed by the fraction of red and blue galaxies from \citet{2007MNRAS.380..585C} in the same redshift range.

%
\begin{table*}
\label{results_morpho}      
\centering                          
\begin{tabular}{l l| r@{}l r@{}l r@{}l | r@{}l r@{}l r@{}l}        
\hline \hline \multicolumn{2}{c|}{}  & \multicolumn{6}{c|}{Local} & \multicolumn{6}{c}{Distant}   \\ \hline 	
\multicolumn{2}{l|}{Type} &\multicolumn{2}{c}{Total ($\%$)} &\multicolumn{2}{c}{Quiescent ($\%$)}&\multicolumn{2}{c|}{Starburst ($\%$)}&\multicolumn{2}{c}{Total($\%$)} &\multicolumn{2}{c}{Quiescent ($\%$)}&\multicolumn{2}{c}{Starburst ($\%$)} \\ \hline
\multicolumn{2}{l|}{\bf{E}}   & 3$\pm$1&      & 3$\pm$2&   &  0$\pm$0&   &   4$\pm$1& & 11$\pm$&3  &  0$\pm$0&    \\ 
\multicolumn{2}{l|}{\bf{S0}}  & 15$\pm$&4     & 14$\pm$&4  & 20$\pm$&10  &  13$\pm$&2 & 33$\pm$&6  &  0$\pm$0&   \\ 
\multicolumn{2}{l|}{\bf{Spiral}}  & 72$\pm$&8 & 76$\pm$&10 & 55$\pm$&17  &  31$\pm$&7 & 31$\pm$&6  & 31$\pm$&8  	\\ 
\multicolumn{2}{l|}{\bf{Peculiar$^{\mathrm{a}}$:}} & 10$\pm$&3 & 7$\pm$3& & 25$\pm$&11  &  52$\pm$&9 & 25$\pm$&5  & 69$\pm$&12   \\

 &P/Irr & 4&$\pm$2  & 2&$\pm$1 &1&5$\pm$9 & 2&6$\pm$7 &  2&1$\pm$5  & 2&9$\pm$8   \\ 
 &P/Tad & 0&$\pm$0  & 0&$\pm$0 & 0&$\pm$0 &   6&$\pm$3  & 0&$\pm$0  & 1&0$\pm$5   \\ 
 &P/Mer & 4&$\pm$2  & 4&$\pm$2 & 5&$\pm$5 &  2&0$\pm$6 &  4&$\pm$2  & 3&0$\pm$8   \\ 
 &P/C   & 2&$\pm$1  & 1&$\pm$1 & 5&$\pm$5 &   0&$\pm$0  & 0&$\pm$0  &  0&$\pm$0    \\ \hline \hline

\end{tabular}
\caption{Fraction of the different morphological types for local and distant samples. $^{\mathrm{a}}$ Fractions in the peculiar class are simply addition of the fraction of peculiar subclasses.}

\end{table*}

%

\section{Building up the past Hubble sequence}

\subsection{Morphological evolution by comparing distant and local galaxies}

The results shown in Table 3 are in a very good agreement with those of \citet{2002PASP..114..797V} who found that the fraction of peculiar galaxies increases from 12\% to 46\% from z=0 to z=0.6-0.8. A simple examination of Table 3 reveals that during the past 6 Gyrs:
\begin{itemize}
\item the fraction of early type galaxies (E/S0) has not evolved, suggesting this population was mostly in place at that epoch, consistently with conclusions of other studies \citep{2007ApJ...669..947J,2006AJ....131.1288B};
\item  the fraction of regular spiral galaxies has increased by a factor 2.3 and this result supersedes former  claims \citep[e.g.,][]{1998ApJ...500...75L} that were based on much lower data quality;
\item the fraction of peculiar galaxies has decreased by a factor 5, and this represents a very strong evolution, as half of the galaxy population consisted of peculiar galaxies 6 Gyrs ago.
\end{itemize}

At first glance, the number evolution of spirals is balanced by the number evolution of peculiar galaxies. Because peculiar galaxies generally show anomalous kinematics \citep{2008A&A...484..159N}, this suggests that peculiar galaxies have to be transformed in one way or another into rotating spirals that we observe in large numbers today. Such a transformation has to be understood and reported by the different models of galaxy evolution. However, before taking that step forward, we need to verify whether this representation of the past Hubble sequence can be causally linked to present-day galaxies.

\subsection{Have we established a past Hubble sequence?}

In this paper we have tried to keep a homogeneous methodology of classifying both distant and local galaxies on the basis of a similar observational set-up. The only difference that could bias our result is the greater optical depth -by 0.08 to 1.0 mag- reached for distant galaxies when compared to local galaxies. To verify whether this might affect our results in Table 3, we have reproduced similar conditions at $z$=0.65 than at $z$=0 by degrading the signal to noise of GOODS observations. It results that small variations of optical depth cannot affect our results in Table 3: all the galaxies which have been classified as peculiar remain in that category. This is not surprising because SDSS imaging is deep enough to provide a fairly robust classification of galaxy morphology \citep[e.g.,][]{2004AJ....127.2511N,2007AJ....134..579F}. Figure 5 shows the two Hubble sequences that have been constructed using both local and distant samples.

   \begin{figure*}
  \begin{center}
  \begin{tabular}{c}
   \includegraphics[width=15cm]{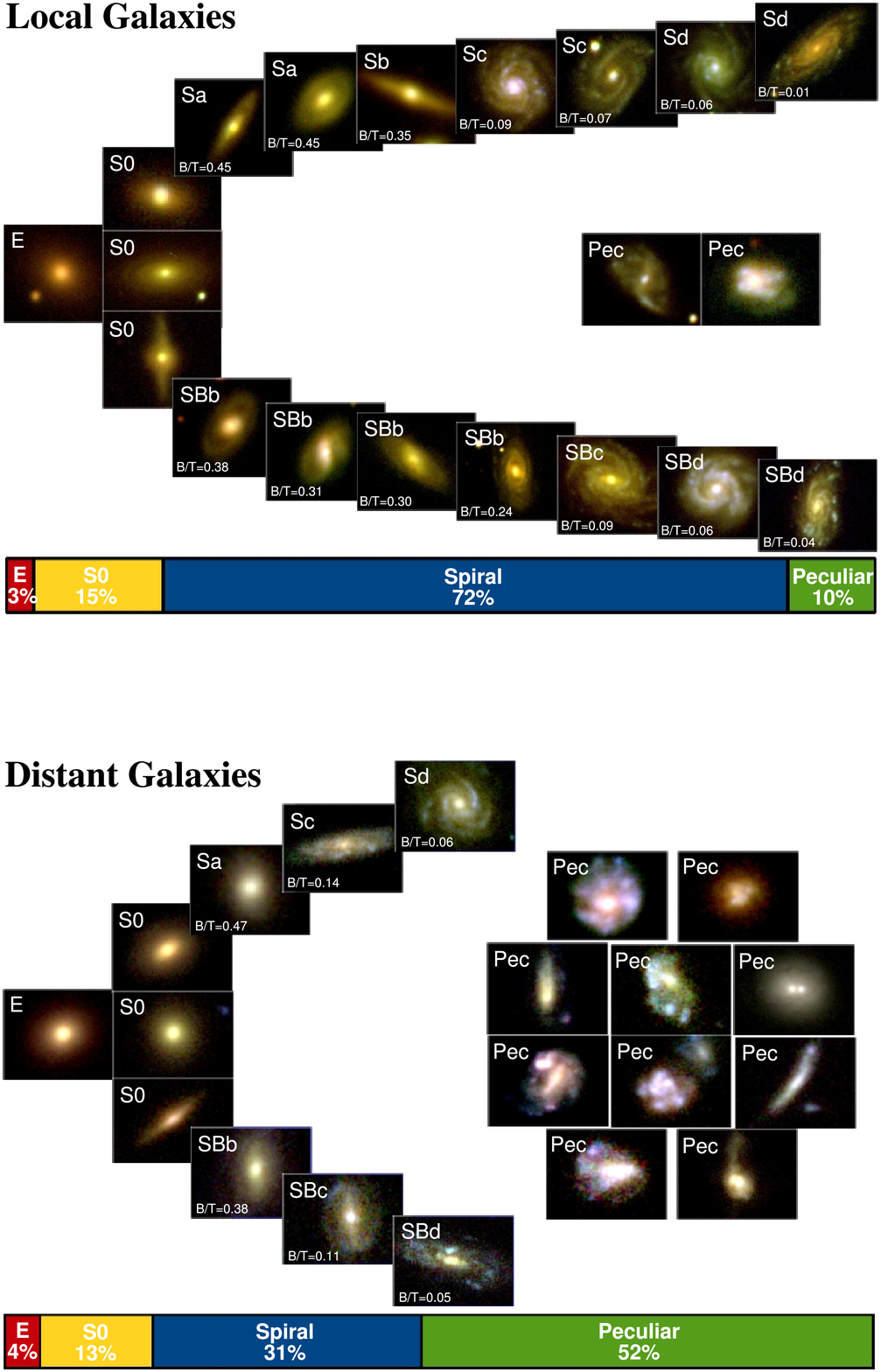} \\
   \end{tabular}
   \caption{Present-day Hubble sequence derived from the local sample and past Hubble sequence derived from the distant sample. Each stamp represents approximately 5\% of the galaxy population. 
            }
   \label{FigHubSeq}
   \end{center}
    \end{figure*}
%

The main drawbacks in establishing such a sequence are, however, related to our fundamental ignorance of the various effects on galaxy morphology during the past 6 Gyrs. As discussed in Sect. 1, the possible effects on either the number density of galaxies or the validity of a common absolute magnitude limit include galaxy merging and stellar population evolution. Their possible impacts are examined in Appendix A. It turns out that the past Hubble sequence displayed in Fig. ~5 provides a reasonable snapshot of the past history of galaxies populating the Hubble sequence today. The link between the two Hubble sequences is marginally affected by very recent mergers or by stellar population evolution. Some very specific star formation histories could, however, complicate the link: for example a galaxy forming all its stars at $z$=1 in a so-called monolithic collapse would show an evolution of its  $M_{stellar}/L_{K}$ ratio by about 0.4 dex according to population synthesis model. Such objects, if they exist, would be found in slightly greater numbers in the past Hubble sequence than in the present-day one. Besides this, the proposed link between past and present sequence is also well adapted to various models of galaxy evolution \citep{2003MNRAS.344.1000B}. For example, let us consider a secular evolution model in which a galaxy gradually accretes gas from the intergalactic medium. Observations \citep{2005A&A...430..115H,2005ApJ...625...23B} suggest a stellar mass increase by 0.3 dex since $z$=1, or 0.15-0.2 dex since $z$=0.65. By selecting local and distant galaxies through  $M_{J}(AB)$$< -$20.3, one would catch both progenitor and descendant because the observed evolution of the $M_{stellar}/L_{K}$ ratio is 0.15 dex.

\subsection{A few results and possible applications}

There are a few results displayed in Table 3 and in Fig.~5 that deserve some comment, as they appear to be unexpected:

\begin{enumerate}
\item The fraction of 10\% of peculiar galaxies in the local Hubble sequence is not negligible, given that these galaxies have stellar masses well above $10^{10}$ $M_{\odot}$. Indeed, \citet{2007AJ....134..579F} find only 1$\%$ of galaxies with peculiar morphologies. This may be related to the not using the color information and thus not accounting for some galaxies (at a level of $\sim$ 3\% in our sample) having a blue ``bulge" surrounded by a redder disk. A part of this discrepancy (an additional 2-3\%) comes from our additional criterion that the bulge and disk centroids should coincide to classify a galaxy as a spiral. Otherwise, local peculiar galaxies seem to be dominated by mergers or compact galaxies.    

\item We also find quite a high fraction ( 25$\%$ ) of peculiar morphologies in the sample of quiescent and distant galaxies. This is not mainly caused by the effects described above. This result slightly affect the conclusions of \citet{2008A&A...484..159N} and \citet{2005A&A...430..115H,2009arXiv0903.3962H}, who assumed that besides E/S0, all quiescent galaxies have to be spirals. This simply reduces further the fraction of regular, rotating disks in the past Hubble diagram whose fraction is 2.3 times lower than in the present-day galaxies.
\end{enumerate}

Figure 5 presents an accurate representation of Hubble sequences at two different epochs for which galaxy morphologies have been studied in an homogeneous way and with a minimum of selection effects. This can also be used to test some scaling relations between fundamental parameters. For example, this has been done by \citet{2006A&A...455..107F} and \citet{2008A&A...484..173P,2009arXiv0903.3961P}, who show that the main evolution of the Tully-Ficher relation is resulting from a much larger scatter from peculiar galaxies. Figure 2 demonstrates that this strong evolution discovered by IMAGES is unaffected by selection biases.

Similarly, one can test the scaling relations of early type galaxies. For example, it has been shown \citep[][and references therein]{1998AJ....116.1591P,2003AJ....125.1849B} that bulge luminosity correlates with the bulge effective radius, which is indeed one projection of the fundamental plane of elliptical galaxies. Figure 6 presents the absolute r-band magnitude of the bulge against the bulge radius for the two subsamples of E/S0 galaxies. Although the number of objects is quite small, there is a correlation between the luminosity and bulge radius. Using the past Hubble sequence, we may verify whether it evolves or not. Figure 6 (bottom panel) shows that no evolution is measured, when accounting for passive evolution of the luminosity (about 0.59 mag), for stellar populations assumed to be formed $\sim$ 12 Gyr ago. 

\begin{figure}[htp]
  \begin{center}
  \includegraphics[height=11cm]{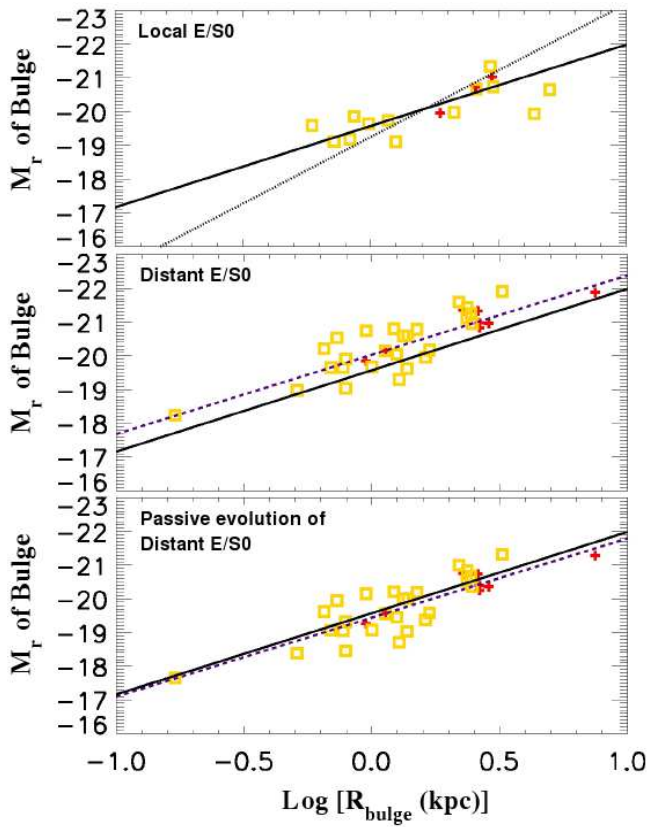} 
  \caption[example] {r absolute magnitude of the bulge vs bulge radius of E (red crosses) and of S0 (yellow squares) galaxies. {\it Top:} Local sample for which the solid line represents the best fit, while the dotted line comes from the \citet{2003AJ....125.1849B} study of elliptical galaxies. {\it Middle:} Same for distant E and S0 galaxies with the dashed line showing the best fit of the relation. There is approxiamtely 0.5 magnitude between the local (solid line) and distant best-fit of the $M_{r}$-$R_{bulge}$ relation. {\it Bottom: }  Same as above but after correcting for a passive evolution of the stellar population. We assume a single star formation burst, 12 Gyrs ago. Other star formation histories could be considered, but this indicates that the $M_{r}$-$R_{bulge}$ is unlikely to be a strongly evolving relationship.}  
  \label{MRvsRB}
  \end{center}
  \end{figure}

\section{Conclusion}

We have established a first approximation of what would be the progenitors, 6 Gyrs ago, of the galaxies of the present-day Hubble sequence. Using a simple and single criterion for selecting both distant and local galaxies, we showed that one may link them together quite robustly. There are certainly uncertainties caused by further, very recent merging events, as well as by various star formation histories, but they are not likely to affect more than 10\% of the galaxies that may be linked together. This is lower than the error due to Poisson number statistics in samples that slightly pass 100 galaxies. Our method classifying galaxy morphologies is similar to that adopted for local galaxies. It uses a semi-automatic decision tree that is applied in a consistent way for both distant and local galaxies. The errors due to misclassification are thus unlikely to be more than a few percent. This leads to:
\begin{enumerate}
\item  E/S0 galaxy populations show no evidence of number evolution during the past 6 Gyrs; 
\item Slightly more than half of the distant galaxies have peculiar morphologies, which is likely associated to anomalous kinematics according to \citet{2008A&A...484..159N};
\item The fraction of regular spiral was 2.3 times lower 6 Gyrs ago than in the present epoch;
\item Almost all the evolution is caused by the transformation of galaxies with peculiar morphologies into regular spiral galaxies at present epoch.
\end{enumerate}

The past Hubble sequence that is established in this study could be a useful tool for further study of galaxy evolution or transformation. It can be used to derive the evolution of fundamental planes for both spirals and bulges. It is, however, unclear whether this method can be extended to longer look-back times, at epochs during which mergers could be a predominant mechanism to form galaxies and their stellar populations. There is, however, a need to improve the statistics, since this paper is based on numbers slightly over 100 for both distant and local samples, i.e., fewer galaxies in the different morphological types. Cosmic variance effects also have to be investigated since they could easily affect the GOODS South field \citep[see][and references therein]{2007A&A...465.1099R}. Further selection of SDSS galaxies and galaxies from GOODS South and North or from deep galaxy surveys with similar depth are feasible. More than half of the galaxies in the past Hubble sequence have peculiar morphologies. Their transformation to regular spirals in the present-day Hubble sequence is a big challenge that should be addressed by current scenarios of galaxy evolution and formation.

\begin{acknowledgements}
      This work is supported by a PhD scholarship from IFARHU-SENACYT (Panama), and grants from Technological University of Panama and the Region {\^I}le-de-France.
\end{acknowledgements}

\appendix

\section{Possible biases and selection effects in linking the two Hubble sequences}

\subsection{Effects related to merging}

Merging naturally affects the evolutionary link between distant and present-day galaxies, simply by a radical suppression of one of the merging galaxies (see bottom of Fig. A.1). Another and more subtle effect would be caused by two galaxies below our limit (e.g. with $M_{J}(AB)$ $> -$20.3) merging and their by-product reaching our limit ($M_{J}(AB)$ $\le$-20.3) at z=0 (see top of Fig. A.1).

\begin{figure}[htp]
  \begin{center}
  \includegraphics[height=6.5cm]{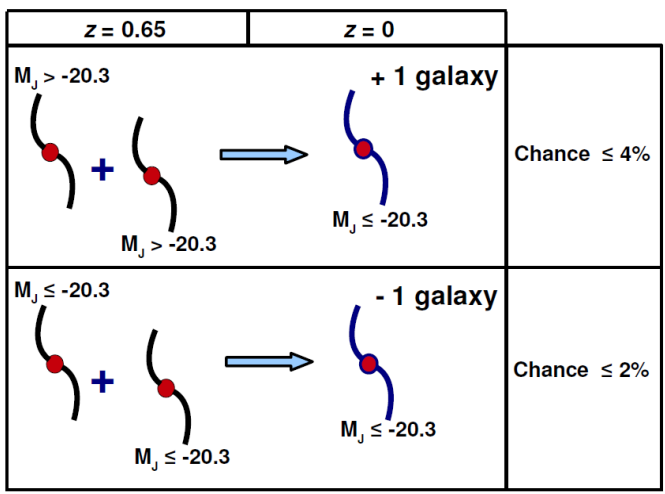} 
  \caption[example] {Two possible merging effects that could affect the link between distant and present-day galaxies. {\it Bottom:} two galaxies with $M_{J}(AB)$ $\le -$20.3 merge after z=0.65 suppressing one galaxy with $M_{J}(AB)$ $\le$-20.3 at z=0.0; {\it Top:} two galaxies with $M_{J}(AB)$ $> -$20.3 merge after z=0.65 giving one galaxy with $M_{J}(AB)$ $\le$-20.3 at z=0.0.}  
  \label{Appendix1}
  \end{center}
  \end{figure}

To evaluate the fraction of z=0.65 galaxies that will merge during the last 6 Gyrs, we may consider the two studies providing the largest and the smallest evolution of the merger rates. Assuming an evolution rate as $(1 + z)^{2.7}$ \citep{2000MNRAS.311..565L}, one can derive that 29\% of $z$=0.65 have to experience a major merger during the last 6 Gyrs. Similarly, \citet{2006ApJ...636..592L} estimate that between 33\% and 66\% of $L_{B}$ $\ge$ $0.4\times$ $L_{B}$ galaxies had a major merger since $z$=1.1, with an evolution rate of $(1 + z)^{1.12}$. Combining the above estimates implies that 15 to 30\% of $z$=0.65 galaxies could experience a major merger later on. This number is not very accurate and depends on assumptions about the selection filter, as well as on the expected time $\tau$ for a galaxy pair to merge. \citet{2008ApJ...681.1089R} have estimated the merger rate using a quite similar method than ours to pre-select galaxy pairs, i.e., within $M_{J}(AB)$ $< -$19.5 galaxies selected in the GOODS field. They do find an evolution following 0.014$\times$$(1 + z)^{2.7}$. For $\tau$=0.35 and 0.5 Gyr, one calculates that 21 and 15\% is the fraction of $z$=0.65 galaxies that are expected to experiment a major merger before reaching $z$=0, respectively. \\

Major mergers may dramatically affect the appearance and dynamical properties of galaxies. Here we are interested in linking distant progenitors with their local descendants. In the following we assume that the actual limit in linking past to present Hubble sequence is realized when a galaxy is simply suppressed by being accreted by a more massive galaxy. \citet{2009arXiv0903.3962H}, assuming that all objects could be modeled as a merger or a merger remnant, have derived the corresponding merger mass ratio (see their Fig. 6, bottom). The probability for an $M_{J}(AB)$ $< -$20.3 to encounter a more massive galaxy is about 7\%. Such a value proves the scarcing of events involving a galaxy more massive than the observed one, since those are rarer due to the exponential drop of the mass function towards the massive end. Thus the chance for a galaxy to be suppressed in the local Hubble sequence by a very recent merger, would be 0.21x0.07, hence less than 2\%.

We also have to account for the possibility that two $M_{J}(AB)$ $> -$20.3 galaxies could merge during the past 6 Gyrs and reach the local Hubble sequence of $M_{J}(AB)$ $< -$20.3 galaxies (see top of Fig. A.1.). The probability of such events is similarly small. To illustrate this, let us assume two progenitors having the same J luminosity: only merging galaxies with $M_{J}(AB)$ $< -$19.6 can be progenitors of an $M_{J}(AB)$ $< -$20.3 galaxy at z=0. From Fig. 2 such galaxies represent less than 20\% of the $M_{J}(AB)$ $< -$20.3 sample. By integrating all possible major mergers in a similar way, it leaves less than 4\% of z=0.65 galaxies that may enter the sample due to a recent major merger. We conclude that very recent mergers are unlikely to affect the link between past and present  Hubble sequences at a level over $\sim$ 5\%.\\

\subsection{Effects related to stellar population evolution}

Our selection is based on the absolute J-band magnitude that is often considered as a proxy for stellar mass. There is also very good correspondence between absolute K-band and J-band magnitudes (see Sect. 2.3). The stellar mass to K-band light ratio is evolving quite significantly with the redshift, which is simply related to the evolution in color or in star formation. For example, applying the \citet{2003ApJS..149..289B} correction to the $M_{stellar}/L_{K}$ ratio, \citet{2008A&A...484..173P} find an average evolution of 0.15 dex when comparing local spiral galaxies to distant emission line galaxies. In other words, by selecting more and more starbursts in the past Universe, we pre-select galaxies with lower stellar masses than those in the local Universe. Such an effect might be non negligible because there are 9\% of distant starbursts that have $-$20.675$<$ $M_{J}(AB)$ $< -$20.3. Interestingly a part of this bias is compensated for when considering the baryonic mass rather than the stellar mass (see Fig. A.2). \citet{2009arXiv0903.3962H} find that distant starbursts have a gas fraction averaging 31\%. This is based on assuming the Kennicutt law that links surface density of star formation rate to that of gas mass. Considering baryonic mass\footnote{We asssume the gas fraction to be 10\% for local galaxies.}, we do find that the larger amount of gas in distant starbursts add a shift of -0.16 dex on to $M_{baryonic}/L_{K}$ ratio, thus compensating quite well for the stellar population evolution. Of course this does not mean that, for any evolutionary scheme, such a compensation may be realized as, for example, \citet{2006ApJ...652...85M} claim for a longer evolution of the $M_{stellar}/L_{K}$ ratio.\\

\begin{figure*}
  \begin{center}
  \includegraphics[height=10.cm]{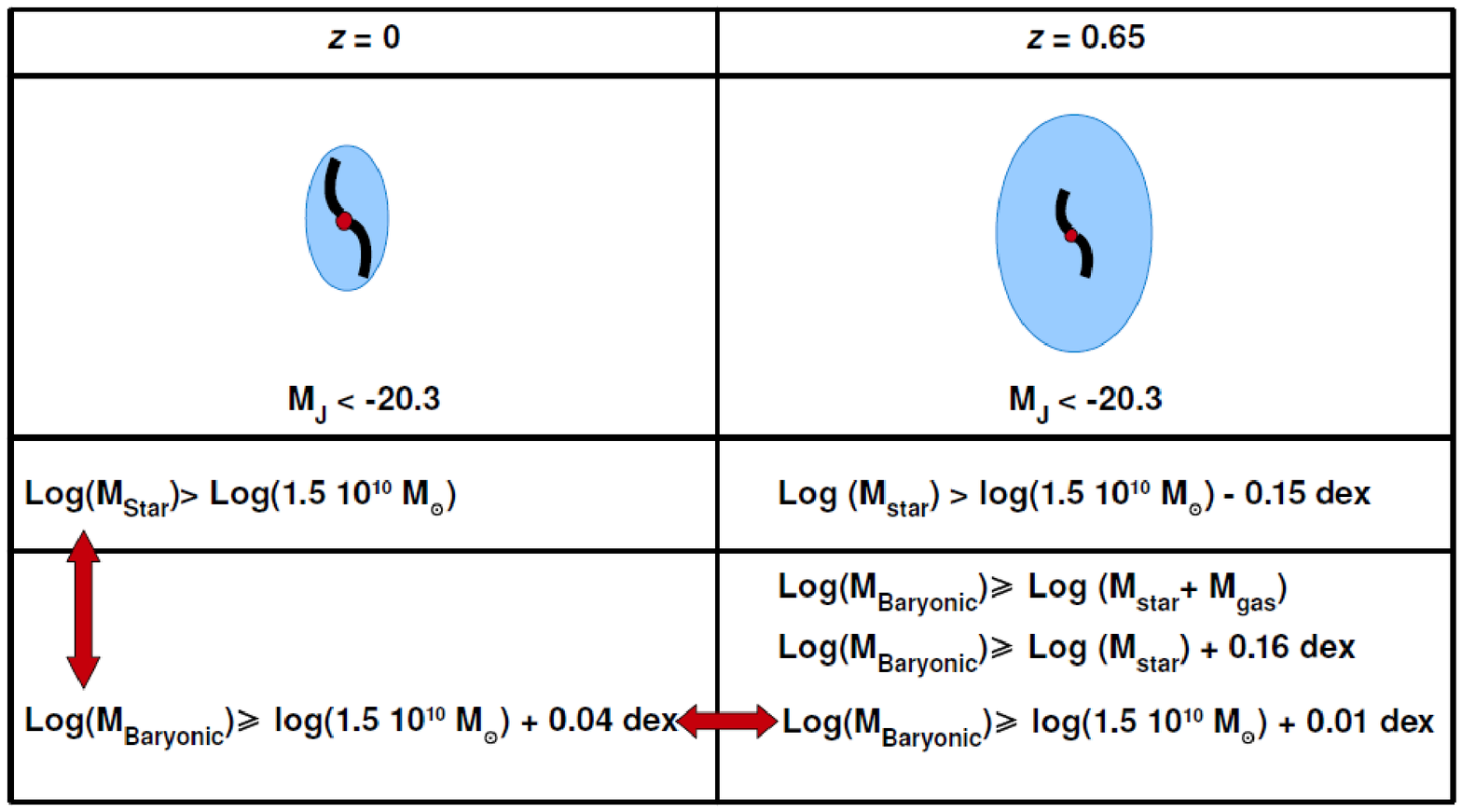} 
  \caption[example] {Stellar and baryonic mass of a galaxy . {\it Left:} galaxy with $M_{J}(AB)$ $< -$20.3 in the local Universe; {\it Right:} galaxy with $M_{J}(AB)$ $< -$20.3 at z=0.65.}  
  \label{Appendix1}
  \end{center}
  \end{figure*}

\end{document}